\documentclass[aps,twocolumn,showpacs,nofootinbib]{revtex4}
\usepackage{graphicx}
\newcommand{\beq}{\begin{equation}}
\newcommand{\eeq}{\end{equation}}
\newcommand{\beqa}{\begin{eqnarray}}
\newcommand{\eeqa}{\end{eqnarray}}

\def\opone{\leavevmode\hbox{\small1\normalsize\kern-.33em1}}

\begin{document}

\title{The Free Will Theorem, Stochastic Quantum Dynamics and True Becoming in Relativistic Quantum Physics}

\author{Nicolas Gisin \\
\it \small   Group of Applied Physics, University of Geneva, 1211 Geneva 4,    Switzerland}

\date{\small \today}

\begin{abstract}
In Bell inequality tests, the evolution of the wavefunction is not covariant, i.e. not invariant under velocity boost that change the time ordering of events, but the laws that govern the probability distribution of possible results are. In this note I investigate what this could mean and whether there could be some covariant "real quantum stuff". This clarifies the implication of the Free Will Theorem and of relativistic spontaneous localization models based on the flash ontology (rGRWf). Some implications for the concept of time(s) are spelled out.
\end{abstract}

\maketitle

The Free Will Theorem\cite{FWT} has attracted quite a lot of attention. This is due in part to provocative terminology, but also to the claim that no deterministic nor stochastic dynamics could ever be compatible with highly plausible assumptions. Without going into (important) details, it will suffice here to recall that the authors of the Free Will Theorem assume the possibility for experimentalists to freely (independently) chose measurement settings and that quantum theory and (special) relativity make correct predictions in already well tested situations, including the violation of Bell's inequality. For deterministic dynamics, this is not new and won't interest us in this note. But for stochastic dynamics the situation is interesting. Indeed, the claim in \cite{FWT} contradicts the explicit model presented by Tumulka in \cite{rGRWf}. The aim of this note is to pin down that the origin of the confusion \cite{controverse} lies in very different understanding of what a "covariant quantum process" is; we shall see that, in some sense, both \cite{FWT} and \cite{rGRWf} are correct.

Quantum measurements applied to systems composed of several distant subsystems, as those used in Bell inequality tests, are at odds with special relativity. Indeed, quantum measurements "collapse" the wavefunction of the system in a non-covariant way. This is true even if one doesn't strictly apply the projection postulate, as long as one admits that (at least some) measurements have definite classical results secured in a finite time. Consequently, the usual wavefunction (or equivalently the state vector) is not a covariant object. This led many authors to conclude that only the probabilities that appear in quantum physics can be described in a covariant way, not the state (see e.g. \cite{AharonovAlbert}).

But, what could this seemingly well accepted "way around the quantum-relativity tension" mean? Should one conclude that the real stuff in quantum physics is not the state, but the probabilities? Or in more dramatic words, that the real stuff are the probabilities, not the probability amplitudes? In this little note I would like to plunge into {\it quantum ontology} and ask what is the real stuff in quantum physics and what are these {\it covariant quantum probabilities}.

Let me start my investigation with the standard example of two qubits, e.g. two 2-level atoms, located at a large distance from each other (large enough that one can easily perform space-like separated measurements on them). First, let's look at the situation from a reference frame in which Alice, who controls the first qubit, performs her measurement first, see Fig. 1. The probability of her outcome can easily be computed and the process simulated on a classical computer. For the purpose of such a simulation one needs a random number generator; one may use a quantum random number generator, but this is not necessary. Actually, one could merely fetch some number from a file that contains random numbers produced and saved to a hard disk a long time ago. Hence, Alice's probabilities can be thought of and simulated as if they were mere epistemic probabilities, that is as if the actual results were determined by some classical variable (the numbers stored on the computer's hard disk). Let me emphasize that all classical probabilities can be thought of and simulated in such a way. Denoting $\lambda$ the classical variables stored on the hard disk, Alice's result $\alpha$ is a function of her measurement settings $\vec a$ and of $\lambda$: $\alpha=F_{AB}(\vec a, \lambda)$, where $F_{AB}$ reminds us that this is the "first measurement" in the time order A-B. Next, consider Bob's measurement, in the same reference frame. His result $\beta$ is also probabilistic, but, since he is second to measure, his result may depends also on Alice's measurement setting and outcome. In order to simulate Bob's outcome one may use the same random number fetched from the hard disk: $\beta=S_{AB}(\vec a, \vec b, \lambda)$, where $S_{AB}$ stands for "second". Note that one could think of $\lambda$ as a nonlocal variable \cite{NLvariable}, but I am more thinking of it as a variable used to simulate such experiments on one (local) classical computer.

That the above sketched simulation works, i.e. reproduced all probabilities and correlations of real experiments, should be clear. Indeed, this is standard quantum mechanics and, again, this is how any stochastic process can be simulated. But now, let's look at the experiment from another reference frame, one in which Bob is first and Alice second, see Fig. 1. By symmetry it is clear that this situation can equally be simulated: $\beta=F_{BA}(\vec b, \lambda)$ and $\alpha=S_{BA}(\vec b, \vec a, \lambda)$ with possibly different functions $F_{BA}\neq F_{AB}$, $S_{BA}\neq S_{AB}$. Note that the same file of $\lambda$'s can be used to equally well simulate both cases, the one in which Alice is first and the one in which she is second. The variable $\lambda$ is arbitrarily large and the functions $F_{AB}$ and $F_{BA}$ may access different parts of $\lambda$.

But now, if the probabilities are the real stuff, as hypothesized above, then there should be a file of appropriate $\lambda$'s, that is, of random variables of the appropriate size and structure, and 4 different functions $F_{AB},F_{BA},S_{AB},S_{BA}$ such that the results don't depend on the reference frame, i.e. don't depend on the chronology, or, in other words, such that, loosely speaking, the "probabilities are covariant":
\beqa
\alpha=F_{AB}(\vec a, \lambda)=S_{BA}(\vec b, \vec a, \lambda) \label{alpha} \\
\beta=S_{AB}(\vec a, \vec b, \lambda)=F_{BA}(\vec b, \lambda) \label{beta}
\eeqa
for all $\vec a, \vec b, \lambda$. But no functions $F_{AB},S_{AB},F_{BA},S_{BA}$ satisfying (\ref{alpha}) and (\ref{beta}) exists. Indeed, (\ref{beta}) implies that $S_{AB}$ is independent of $\vec a$ which would turn $\lambda$ into a local variable. And it is well known that local variables can't simulate all quantum correlations, in particular can't simulate those violating Bell's inequality.

\begin{figure}
\includegraphics[width=9cm]{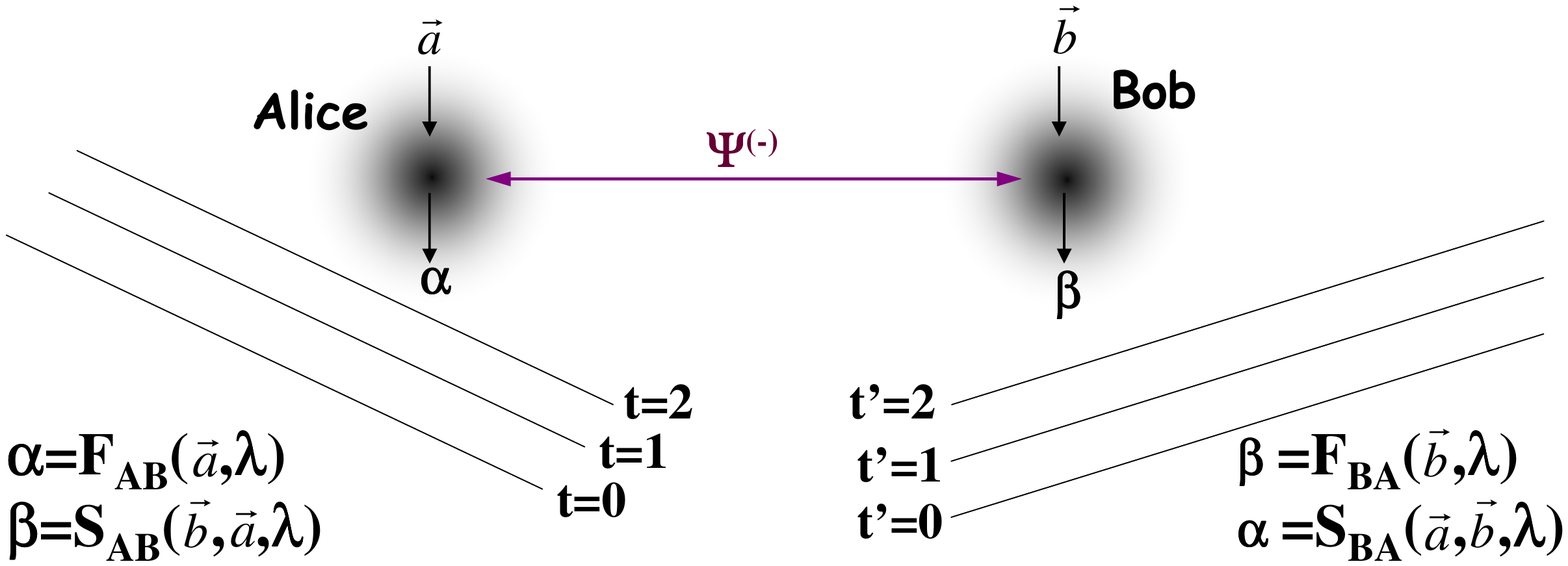}
\caption{\it Standard configuration for tests of Bell inequality as seen from two different reference frames. In the first frame, Alice performs her measurement before Bob and her result is determined by the function $F_{AB}$; in the other frame she is second and her result is given by the function $S_{BA}$, as explained in the text.}
\end{figure}

Accordingly, not only can the quantum state not be described in a covariant way, but neither can the probabilities. So, what remains?

Here one needs to be more careful and precise with what one means by {\bf "covariant probabilities"}: one should distinguish probability {\bf distributions} from their {\bf realizations}. Intuitively, a probability distribution is like a cloud of potential events that may occur with some probabilities (e.g. they have an intrinsic propensity to realize themselves), while a realization is a single event that happens to became actual. This is illustrated in Fig. 2 where the star near the center distinguishes the actual event among the cloud of possibilities.

Accordingly, the probability distributions remain covariant (the sets of possible outcomes is covariant), but not their realizations. If one looks at the same experiment from two different reference frames, one can assign compatible probability distributions (this is what standard quantum physics does), but there is no way to assign unique realizations of these probability distributions. The consequence is that one should not think of these probabilities as usual; in particular one can't simulate these probabilities on a classical computer, or, more precisely, if one simulates them, the simulation can be faithful only for a given chronology.

So, what, then, is real? The probability distributions or the actual realizations? The probability distributions are covariant, but don't describe the actual world. The realizations describe the world we see around us, but, necessarily, in a non-covariant way. In short: Only the cloud of potentialities is covariant, the actual events aren't. Hence, in some sense, the open future is covariant, but the past is not.

\begin{figure}
\includegraphics[width=9cm]{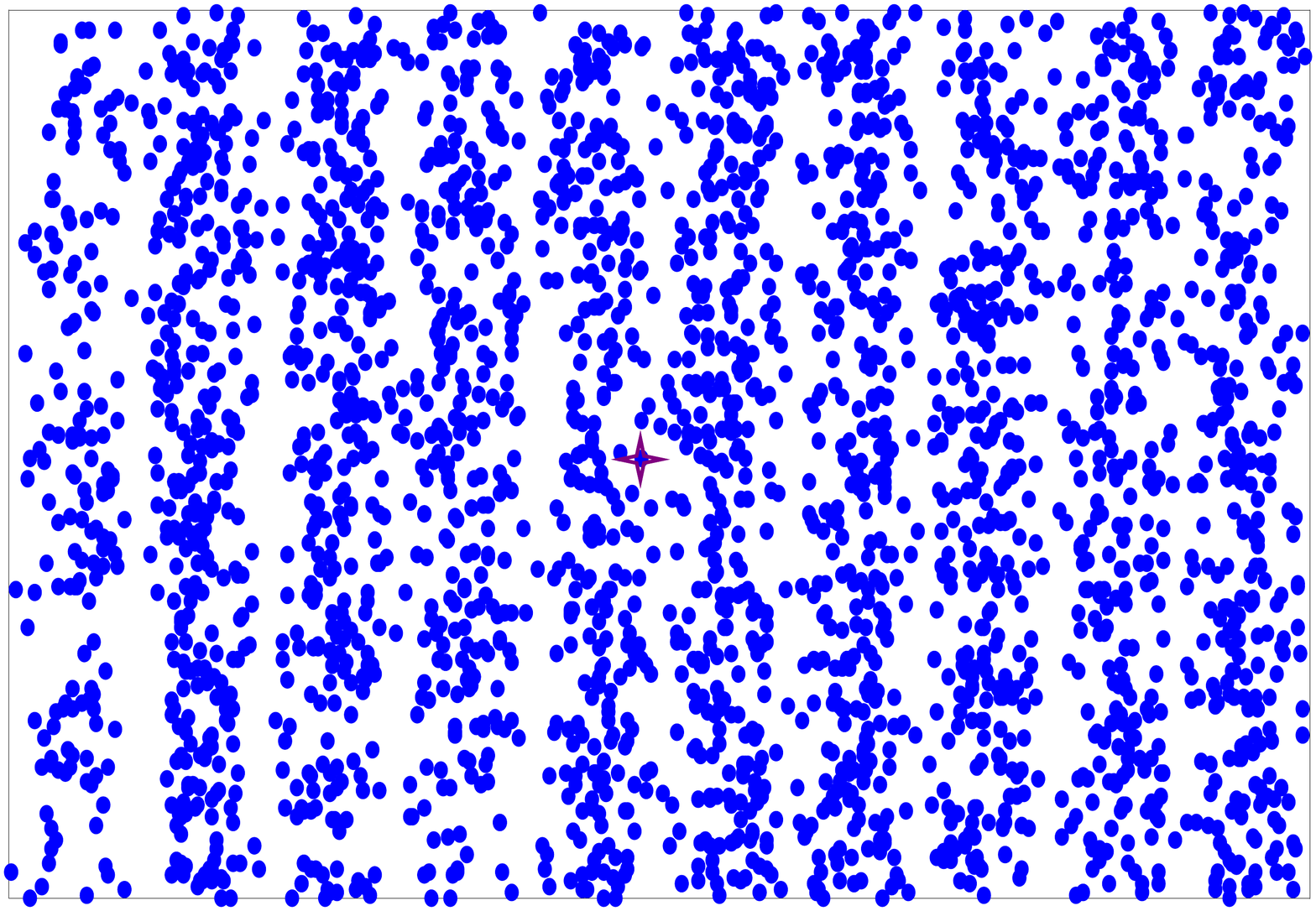}
\caption{\it A cloud of possible future events, as in standard interferometric experiments. The cloud suggests a probability distribution. The star near the center highlights the event that actually happened in the experiment. The cloud of potential events is covariant, but the actual event is not.}
\end{figure}

Note the implication for the concept of time. Quantum events are not mere functions of variables in space-time, but true creations: time does not merely unfold, true becoming is at work. The accumulation of creative events is the fabric of time \cite{TwoTimes}.

But does all this have any practical implications? Yes, of course. Let's come back to the Free Will Theorem (FWT) and to Tumulka's relativistic stochastic spontaneous localization model. In \cite{FWT} the authors of the FWT correctly recall that the free choice of settings in a Bell inequality test, the violation of the inequality and relativistic time-order invariance are incompatible with any deterministic evolution of the quantum stuff (quantum state, quantum particle, name it as you like). But then, the authors claim that their result extends straightforwardly to the non-deterministic case, arguing that "it will plainly make no difference to let them (the random numbers $\lambda$) be given in advance"\cite{FWT}. As we have seen in this note, this claim is correct if one insists that the realizations should be time-order invariant, but not if one is satisfied with time-order invariant (i.e. covariant) probability distributions. Hence, the FWT states that the quantum events are not functions of space-time, a result one can find under many different forms in the literature, and, more interestingly, states as a consequence that the quantum events must enjoy some sort of "freedom", i.e. in my words correspond to true becoming. Let me add that, although the FWT isn't really new and the proof given unnecessarily complicated, I am very sympathetic with the spirit of the FWT: quantum events are not merely the realization of usual probability distributions, but must be thought of as true acts of creations (true becoming).

Let me turn to Tumulka's nice model and to the controversy he and co-workers raised against the FWT. He named his model  rGWRf for "relativistic GRW model with the flash ontology" \cite{rGRWf}. In this model, the quantum stuff are the locations of the spontaneous localization (GRW hits), that he calls flashes. The probability distributions of possible future flashes is time-order invariant, i.e. covariant. But the set of actual past flashes are not (and couldn't be). Hence, the rGRWf model is as covariant as possible: the laws governing rGRWf are covariant. By abandoning the usual {\it wavefunction ontology} for the {\it flash ontology}, Tumulka gave us the best model we have today \cite{ToyModel}. Tumulka and co-authors correctly insist that rGRWf is as covariant as possible, and the FWT correctly stresses that it is not more covariant than possible, as this little note hopefully clarifies.


\small

\section*{Acknowledgment} This work profited from comments by Valerio Scarani, Stefano Pironio, Simon Kochen and Roderich Tumulka.


\begin{thebibliography}{99}
\bibitem{ToyModel} Though Tumulka's model is still only a toy model since it doesn't allow for interactions. Note that this doesn't prevent one to apply it to interesting situations, as those occuring in Bell inequality tests.
\end{thebibliography}
\end{document}